\documentstyle[12pt]{article}           

\def\preprint{IFT-01-2002}       
\def\finished{January 2002}
\def\archive {hep-th/0102XXX}           

\def\title{Covariantization and canonical quantization.}

\long\def\abstract{
In the light-cone gauge choice for Abelian and non-Abelian gauge
fields, the vector boson propagator carries in it an additional
``spurious'' or ``unphysical'' pole intrinsic to the choice requiring
a careful mathematical treatment. Research in this field over the
years has shown us that mathematical consistency only is not enough to
guarantee physically meaningful results. Whatever the prescription
invoked to handle such an object, it has to preserve causality in the
process. On the other hand the covariantization technique is a well
suited one to tackle gauge dependent poles in the Feynman integrals,
dispensing the use of {\em ad hoc} prescriptions. In this work we show
that the covariantization technique in the light-cone gauge is a
direct consequence of the canonical quantization of the theory.
}

\input{epsf}

\begin{document}

{\hfill \archive   \vskip -2pt \hfill\preprint }
\vskip 15mm
\centerline{\Large\bf Covariantization and}
\vskip 7.5mm
\centerline{\Large\bf canonical quantization}
\vskip 7.5mm
\centerline{\Large\bf in the light-cone gauge}
\vskip 8mm
\begin{center} \vskip 10mm
Alfredo T. Suzuki{\begingroup\def\thefootnote{a}
                                \footnote{e-mail: suzuki@ift.unesp.br}
                                \addtocounter{footnote}{-1} \endgroup}
and \  \
Ricardo Bent\'{\i}n{\begingroup\def\thefootnote{z}
                                \footnote{e-mail: rbentin@ift.unesp.br}
                                \addtocounter{footnote}{-1} \endgroup}
\\[3mm] Instituto de F\'{\i}sica Te\'orica - UNESP.\\
        Rua Pamplona 145, CEP 01405-900, SP, Brasil.
        
\vfill                  {\bf ABSTRACT } \end{center}    \abstract

\vfill \noindent \preprint\\[5pt] \finished \vspace*{9mm}
\thispagestyle{empty} \newpage
\pagestyle{plain} 

\newpage
\setcounter{page}{1}

\baselineskip1cm

\section{\large Introduction.}
The historical development of light-cone gauge started well back in
the 1940's with the pioneering work of P.A.M.Dirac \cite{dirac}. Since
its d\'ebut into quantum field theory it has known days of both glory
and oblivion for varied reasons. On the one hand it seemed a solid
grounded and more convenient approach to studying quantum fields,
e.g., the only setting where a proof of the finiteness of the $N=4$
supersymmetric Yang-Mills theory could be carried out successfully was
in the light-cone gauge (a facet of its glory) \cite{susylc}. But on
the other hand, manifest Lorentz covariance is lost and non-local
terms sneak into the renormalization program (the other side of the
coin that charges us with a price to pay). \\
Having this very brief historical overview of the light-cone gauge as
our hindsight, we can say that some of the advantages of using it in
quantum field theory has to do with the possibility of decoupling the
ghost fields in the non-Abelian theories, since it is an axial type
gauge, as shown by J. Frenkel \cite{josif}, a property that can
simplify Ward-Takahashi identities \cite{WTI} and problems involving
operator mixing or diagram summation \cite {opmix}.\\
The annoying thing is that actual calculations of Feynman diagrams
involved in any physical process of interest give rise to ``spurious''
poles and these need to be treated carefully. By carefully we mean
that any device or prescription called upon to handle the singularity
needs not only to be mathematically sound but must guarantee no
violation of basic underlying physical principles such as
causality \cite{presc}. Besides, even with causal prescriptions such as the
Mandelstam-Leibbrandt one to handle it, there emerges a lot of
nonlocal terms in the calculation proper that could pose challenges to
the renormalization program \cite{various}.\\
Of course, the machinery developed to handle covariant poles in
propagators is on the road for a longer period of time, and from the
start the procedure there has been such that the methodology sought 
would
preserve unitarity, causality and so on, i.e., the analytical
properties being rightly understood and applied for that case, ensuring
the possibility of Wick rotation. Therefore if we could somehow build
a bridge between the light-cone and the covariant pole, that could
bring some fruition to our understanding of the problem inherent to
the light-cone. This is exactly what has been proposed in \cite{suz}
and the bridging of the gap between the two kinds of singularities was
named ``covariantization''.\\
Covariantization then is a procedure to treat the so called
``spurious'' poles that appear in the light-cone gauge choice
transforming them into covariant-type poles, by the use of a
``dispersion'' relation. This enables us to take advantege of well
known properties belonging to covariant poles in the propagators,
eliminating the need to construct an ``ad hoc'' prescription to handle
especifically the light-cone pole. As far as we know there is only one
other prescriptionless method to handle light-cone integrals: the
negative dimensional integration method (NDIM) which is an altogether
different approach developed later by the turn of the last century
\cite{gente}.\\
The above mentioned covariantization method was applied from an
intuition that the similarity of covariant-type pole generated by the
dispersion relation would work correctly for the light-cone pole thus
covariantized. In order to be formal and consistent, it also has to be
possible to deduce it from the canonical quantization of the
theory, and this is the aim and purpose of the present work. Since
ghosts in a non-Abelian theory decouple in the light-cone gauge and
without loss of generality, it is sufficient for us to consider an
Abelian theory for this purpose.\\
The outline for out paper is as follows: in the next short section we
review and consider the main ideas behind covariantization method
whereas in the following section we consider the canonical
quantization of a model Abelian quantum gauge field constrained in the
light-cone choice. The last section is devoted to the final comments
and conclusions.\\
\section{\large Covariantizaton.}
Here is a brief review of the ``covariantization'' technique which was
proposed by A.T.Suzuki \cite{suz}. The idea is quite simple. In 
light-cone
coordinates, the square of a four-momentum is:
\begin{equation}
  q^2=2q^+q^--\hat{q}^2\;.
\end{equation}
Therefore, as long as {\it $q^-\ne0$} (and this is a key point, as we
shall shortly see), we can write $q^+$ as
\begin{equation}
  q^+=\frac{q^2+\hat{q}^2}{2q^-}
\end{equation}

We note that this dispersion relation {\em almost} guarantees that real
gauge fields for which $q^2=0$ (real photons or real gluons for
example) are transverse; the residual gauge freedom, that is left to
be dealt with so that fields be manifestly transverse comes from the
presence of the $q^-$ in the denominator of the expression above.
This implies that in the light-cone gauge the characteristic pole
becomes
\begin{equation}
  \frac{1}{q^+}=\frac{2q^-}{q^2+\hat{q}^2}\;,\;\;\;\;\;q^-\ne 0\;.
\end{equation}
Note that the mapping between the light-cone pole in $q^+$ and the
covariant-type pole $q^2+\hat{q}^2$ is restricted by the condition
$q^-\ne 0$. This means that, in the usual coordinate notation
$q^-=q^0-q^3\ne 0$, or $q^0\ne q^3$.\\
If we remind ourselves of the pure covariant case, this is equivalent
to the condition $q^0\ne 0$, the zero-mode extracted out from the
range of meaningful frequencies allowed for the quanta: only negative
energy quanta $q^0<0$ propagates into the past and only positive
energy quanta $q^0>0$ propagates into the future. This is the causal
connection. In the light-cone case, this translates into two
distinct physically allowed regions, namely $q^0<q^3$ and $q^0>q^3$
respectively.\\
The important thing here is that the condition $q^-\ne0$ warranties
the causal structure of the covariantization technique since it
eliminates the troublesome $q^-=0$ modes. Elimination of these modes
restores the physically acceptable results as can manifestly be seen
in the causal prescription \cite{pz} for the light-cone gauge.

\section{Canonical quantization.}
As explained earlier, it is sufficient for us to consider here the
Abelian gauge fields. Our procedure is similar to that used by
A. Burnel \cite{burnel} in relating the canonical formalism to the
Mandelstam-Leibbrandt prescription for the light-cone gauge.\\
Let us start with:
\begin{equation}
  {\cal L} = -\frac{1}{4}(F_{\mu\nu})^2
  + h\; (m\cdot \partial)(n\cdot A) + \frac{1}{2}\alpha\; h^2,
\end{equation}
where $h$ is an auxiliary field \cite{por,kugo}, $\alpha$ is a
parameter, $n_\mu$ and $m_\mu$ are the light-cone vectors that
define the gauge (note that we have two null vectors that define the
gauge, even though the usual gauge condition mentions only $n\cdot
A=0$), and the field strength tensor
$$
  F_{\mu\nu}=\partial_\mu A_\nu-\partial_\nu A_\mu.
$$
Using the Euler-Lagrange conditions, we obtain the equations of motion:
\begin{eqnarray*}
  \partial^\mu F_{\mu\nu}-h\;(m\cdot\partial)\; n_\nu &=&0,\\
  (m\cdot\partial)\; (n\cdot A)&=&\alpha \; h.
\end{eqnarray*}
Following the canonical formalism, the equal-time commutation relations 
are:
\begin{eqnarray}
  \label{canonical}
  {[A_\mu(x),\pi^\nu(y)]}_{x_0=y_0}&=&i\delta^\nu_\mu\prod_{i=1}^3\delta(x^i-y^i),\\
  {[A_\mu(x),A_\nu(y)]}_{x_0=y_0}&=&0,\\
  {[\pi^\mu(x),\pi^\nu(y)]}_{x_0=y_0}&=&0,
\end{eqnarray}
where the canonical momenta are
\begin{eqnarray}
  \label{pi}
  \pi^0&=&h\;n_0\;m_0,\\
  \pi^k&=&F^{k0}+h\;m_0\;n^k.
\end{eqnarray}
Using the equal-time relations (\ref{canonical}) and the equations of
motions, it is possible to get the commutation relations for any
time. This was done by Burnel in 1989 \cite{burnel}. That we are
interested in is on momentum space operators, so we will leave
space-time in the usual way, i.e. using the Fourier transforms
\begin{equation}
  \label{fourier}
  A_\mu(x)=\frac{1}{(2\pi)^{2/3}}\int d^4k 
  \theta(k_0){[a_\mu(k)e^{-ik\cdot x}+b_\mu(k)e^{ik\cdot x}]},
\end{equation}
with the following commutation relations for $a_\mu$ and $b_\mu$:
\begin{eqnarray}
  \label{ab}
  \theta(k_0)[a_\mu(k),b_\nu(k')]&=&\delta^4(k-k')\;{\biggl[}-g_{\mu\nu}\delta(k^2)\\
 &-&\alpha k_\mu k_\nu\;\delta({k\cdot n\;k\cdot m})\nonumber \\
 &+&n^2(k\cdot m)^2k_\mu k_\nu\left(\frac{k\cdot n\;k\cdot 
m}{k^2}-\frac{k^2}{k\cdot n\;k\cdot m}\right)\nonumber \\   
&+&(n_\mu k_\nu+n_\nu k_\mu)k\cdot m\left(\frac{\delta(k\cdot n\;k\cdot 
m)}{k^2}+\frac{\delta(k^2)}{k\cdot n\;k\cdot 
m}\right){\biggr]}\nonumber
\end{eqnarray}
with also $[a_\mu(k),a_\nu(k')]=0$ e $[b_\mu(k),b_\nu(k')]=0$.\\
The propagator is defined by
\begin{equation}
  \label{prop}
  \tilde{G}_{\mu\nu}(x)=<0|T[A_\mu(x)A_\nu(0)]|0>
\end{equation}
Using the equations (\ref{fourier}) and (\ref{ab}), we obtain
$$
  \lgroup\Box g^{\mu\nu}-\partial^\mu\partial^\nu
  -n^\mu n^\nu\frac{(m\cdot 
\partial)^2}{\alpha}\rgroup\tilde{G}_{\nu\rho}
  =i\delta^\mu_\rho\delta^4(x),
$$
and inverting this last equation we arrive at
\begin{eqnarray}
  \label{covariant}
  G_{\mu\nu}(k)=\frac{-i}{k^2}{\biggl[}g_{\mu\nu}
  -\frac{n_\mu k_\nu+n_\nu k_\mu}{k\cdot n\;k\cdot m}k\cdot m
  -\frac{\alpha k^2-n^2(k\cdot m)^2}{k\cdot n\;k\cdot m}k_{\mu\nu}
  {\biggr]}\,.
\end{eqnarray}
That for the $\alpha\rightarrow 0$ limit in the light-cone gauge
($n^2=0)$ {\bf goes exactly to the covariantization proposal}. 
\begin{equation}
G_{\mu\nu}(k)=\frac{-i}{k^2}{\biggl[}g_{\mu\nu}
  -\frac{n_\mu k_\nu+n_\nu k_\mu}{k\cdot n\;k\cdot m}k\cdot m
  {\biggr]}\,.
\end{equation}

Note that this preserves the causality of the propagator since the
$k^{-}=k\cdot m$ mode does not vanish nor cannot be canceled against
the $k^{+}\,k^{-}=k\cdot n\,k\cdot m$ bilinear term in the
denominator. This fact is also observed in the definition of the
light-like planar gauge as was done in \cite{planarlcg}.

\section{\large Conclusions.}

In this work we have demonstrated that the canonical quantization of
the theory in the light-cone gauge leads to the covariantization
proposal for the gauge dependent pole. The question that we now ask is
related to the importance of the poles with the bilinear terms $(k\cdot
n\;k\cdot m)^{-1}$ instead of the usual ones $(k\cdot n)^{-1}$. The
raising of such a question becomes natural since that bilinear
structure is the one that preserves causality of the theory. This is
also related to the form of the propagator in the case of the
light-like planar gauge \cite{planarlcg}, where the important fact was
to observe the discrete simmetry between the light-cone vectors
$n_\mu$ and $m_\mu$.

However, here, our starting term for the gauge fixing Lagrangian,
${\cal L}_{\mbox {fix}}$, does not display this kind of discrete
simmetry, since we took a general class III linear gauge. So, what
seems the logical step now is to seek a gauge fixing Lagrangian term
of the form
\[
{\cal L}_{\mbox {\small {fix}}}={\cal L}_{\mbox {\small 
{fix}}}(n,\,m,\,A),
\]
that is, a term with such discrete simmetry built into it. The easiest
way to construct or incorporate this simmetry is to take the factors
of the form $(n\cdot A)(m\cdot A)$ alongside some non-propagating
auxiliary fields. Then, some tentative forms would be terms such as
these:
\begin{itemize}
\item First proposal:
\[
{\cal L}_{\mbox {\small {fix}}} = h\,[(n\cdot A)(m\cdot A)]^{\frac 
{1}{2}}-\frac {1}{2}\,h^2\,,
\]
\item Second proposal:
\[
{\cal L}_{\mbox {\small {fix}}} = h\,(n\cdot A)-\frac {1}{2}\,\frac 
{(n\cdot A)^2}{(n\cdot A)(m\cdot A)}\,h^2\,,
\]
\item Third proposal:
\[
{\cal L}_{\mbox {\small {fix}}} = h\,(m\cdot A)-\frac 
{1}{2}\,\frac{(m\cdot A)^2}{(n\cdot A)(m\cdot A)}\,h^2\,.
\]
\end{itemize}
where for simplicity we have taken the gauge parameter $a=1$.
The first option has the built-in simmetry $n\leftrightarrow m$, but
it is practically prohibited by the very presence of a square root in
it. The second and third options does generate the desired gauge
fixing term ${\cal L}_{\mbox {\small {fix}}}$, but they lack the 
manifest
simmetry $n \leftrightarrow m$. So, is it the best we can arrive at? 
Hopefully not, since upon examining the three proposals above, one notes 
that they have one feature in commom: they make use of just {\em one} 
non-propagating auxiliary field $h$. If we insert a second auxiliary 
field, we can construct a Lagrangian density of the form
\begin{equation}
\label{nm}
{\cal L} = -\frac {1}{4}(F_{\mu\nu})^2+h_1\,(n\cdot A)+h_2\,(m\cdot 
A)-h_1\;h_2
\end{equation}
which, of course, is more complicated, but nonetheless bears the
desired features we want. Just in passing we mention that the
Lagrangian density in (\ref{nm}) yields a propagator of the exactly
same form as found in \cite{planarlcg}. Our interest now is the
canonical quantization of the Lagrangian density (\ref{nm}) whose
result is to appear shortly elsewhere.
  
\vspace{.5cm}

{\bf Acknowledgments:} A.T.S. acknowledges partial funding from CNPq 
and R.B. wishes to thank {\sc fapesp} for 
financial support.


\vspace{.5cm}

\begin{thebibliography}{11}

\bibitem{dirac} P.A.M.Dirac, Rev. Mod. Phys. {\bf 21} (1949) 191
\bibitem{susylc} J.H.Schwarz, Phys. Rep. {\bf 89} (1982) 223; 
S.Mandelstam, Nucl. Phys. {\bf B213} (1983)183; 
L. Brink {\em et al}, Phys. Lett. {\bf 123B} (1983) 323 
\bibitem{josif} J. Frenkel,
  Theories in Algebric Non-Covariant Gauges, World Scientific (1991).
\bibitem{WTI} D.M.Capper, and D.R.T.Jones, Phys. Rev. {\bf D31} (1985) 
3295
\bibitem{opmix} C.B.Thorn, Phys. Rev. {\bf D20} (1979) 1934;
D.J.Pritchard and W.J.Stirling, Nucl. Phys. {\bf B165} (1980) 237
\bibitem{presc} S. Mandelstam, Nucl. Phys. {\bf B213} (1983) 149,\\
  G. Leibbrandt, Can. J. Phys. {\bf 64}, (1986) 606,\\
  B.M.Pimentel, and A.T.Suzuki, Phys. Rev. D {\bf 42} (1990) 2115,
\bibitem{various} A.Bassetto, G.Nardelli, and R.Soldati, {\em 
Yang-Mills Theories in Algebraic Noncovariant Gauges}, World Scientific, 
Singapore (1991);
G.Leibbrandt, {\em Noncovariant Gauges: Quantization of Yang-Mills and 
Chern-Simons Theory in Axial-type Gauges}, World Scientific, Singapore 
(1994)
\bibitem{suz} A.T.Suzuki, Modern Phys. Lett A, {\bf 8} (1993) 2365,
\bibitem{gente} A.T.Suzuki, A.G.M.Schmidt, and R.Bentin, Nucl. Phys. 
{\bf B537} (1999) 549,
\bibitem{pz} B.M. Pimentel and A.T. Suzuki, Phys. Rev. D 42
  (1990) 2115,
\bibitem{burnel} A. Burnel, Phys. Rev. D {\bf 4} (1989) 1221,
\bibitem{por} S. Porkoski, Gauge Fields Theories, 2nd. ed. Cambrigde
  Monographs on Mathematical Physics, (2000),
\bibitem{kugo} T. Kugo and S. Uehara, Nucl. Phys. {\bf B197} (1982) 378,
\bibitem{planarlcg}  A.T.Suzuki and R. Bentin, 
  Modern Phys. Lett A, {\bf 17} (2002) 205.
\end{thebibliography}
\end{document}